\documentclass[pra,amsfonts,twoside,amssymb,superscriptaddress,twocolumn]{revtex4-1}
\usepackage{bbm}
\usepackage{graphicx}
\usepackage{dcolumn}
\usepackage{amsmath}
\usepackage{epsfig}
\usepackage{xcolor}
\usepackage{subfigure}
\usepackage{booktabs}
\usepackage{multirow}
\usepackage{makecell}
\usepackage{booktabs}
\usepackage{mathrsfs}

\usepackage{IEEEtrantools}
\usepackage{float}
\usepackage{amsfonts}
\usepackage{algorithm}
\usepackage{algpseudocode}
\floatname{algorithm}{Algorithm}

\usepackage{hyperref}

\begin{document}

\title{Quantum multi-label $k$-nearest neighbor}
\author{Yilin Shen}
\author{Qin Liao}\email{llqqlq@hnu.edu.cn}
\affiliation{College of Computer Science and Electronic Engineering, Hunan University, Changsha 410082, China}

\date{\today}

\begin{abstract}
Although multi-label $k$-nearest neighbor (ML-$k$NN) is able to effectively solve multi-label learning (MLL) problem with local neighborhood similarity, its time complexity is nearly unacceptable with large-scale datasets. To solve this issue, we propose a novel ML-$k$NN algorithm with quantum computing techniques, which called quantum multi-label $k$-nearest neighbor (QML-$k$NN). In particular, we first accelerate the calculation of the prior probability by taking advantage of quantum phase estimation and Grover's amplitude amplification. Then, a controlled-SWAP test and a quantum $k$-maximal similarity search are used for efficiently identifying the neighbors. Subsequently, a quantum parallel counting circuit (QPCC) is designed to rapidly calculate the posterior probabilities. Experimental results demonstrate that QML-$k$NN is able to significantly reduce the time complexity of solving multi-label problems with performance improvement, achieving a substantial speedup over the classical MLL algorithm.
\end{abstract}

\maketitle

\section{Introduction}
\label{Sec1}
Traditional supervised learning stands as one of the most studied paradigms in machine learning, where each real-world instance is represented by a single feature vector and associated with a single label \cite{07supervised}. However, real-world objects often exhibit inherent polysemy and complexity. For example, an image in scene classification can be assigned to multiple semantic classes such as $sea$ and $sunset$, and a document in text categorization can belong to several predefined topics such as $politics$, $economy$ and $technology$ \cite{10supervise}. The conventional single-label framework is insufficient to capture such rich semantic information, which has motivated the development of multi-label learning (MLL) \cite{14mll-review, 22mll}. The central objective of MLL is to learn a mapping from the feature space to a set of labels, enabling each instance to be associated with a subset of labels. This capability is critical for applications including image annotation \cite{04pattern}, gene functional analysis \cite{20gene}, and text classification \cite{21textclassify, 21textclassify2}.

In general, the effectiveness of MLL largely depends on adequately modeling correlations among labels. For instance, in scene images, $beach$ and $urban$ often co-occur, whereas $fall foliage$ is frequently mutually exclusive with both. Neglecting such correlations can introduce significant prediction bias \cite{23correlation}. To capture label co-occurrence, numerous algorithms based on the principle of local similarity have been developed, such as text categorization \cite{19categorization}, decision trees \cite{15tree}, and multi-label kernel methods \cite{23kernel}. Among these, the multi-label $k$-nearest neighbor (ML-$k$NN) algorithm stands out as an effective lazy learning method \cite{07ml-knn}. It successfully integrates the local neighborhood similarity of $k$NN with the Bayesian rule \cite{67knn}, preserving the instance-based nature of $k$NN while learning label correlations without explicit training. This approach mitigates the challenge of combinatorial label explosion, e.g., capturing the frequent co-occurrence of $politics$ and $economy$ in news articles, and has been shown experimentally to outperform several well-established multi-label algorithms across diverse datasets \cite{07ml-knn}.

Recent research on ML-$k$NN has evolved along the collaborative axes of algorithm optimization, domain adaptation, and technology integration. For example, a rough set-weighted variant named NRFD\_KNN was proposed to improve classification accuracy by identifying uncertain samples in boundary regions and weighting attributes based on their discriminative power \cite{22weight-mlknn}. For multi-label data streams, a weighted ensemble method (WENNML) dynamically selects and weights classifiers to adapt to concept drift \cite{23weighted}. Another study introduced a robust imbalance variant (IRMLSA) that integrates geometric mean-based self-adjusting windows and instance-specific dynamic thresholds to retain minority-label instances in streaming scenarios \cite{24Self-Adjusting}.

Although these studies have improved the performance of ML-$k$NN in various aspects, they have largely overlooked a fundamental computational challenge. When applied to large-scale and high-dimensional datasets, the time complexity of most existing variants becomes nearly unacceptable, with substantial costs incurred at virtually every algorithmic stage, including neighbor search and probability calculation \cite{16dimension}. Consequently, despite its strong predictive capacity, ML-$k$NN remains hampered by significant computational bottlenecks.

Fortunately, recent advances in quantum computing have created novel avenues for overcoming these classical limitations \cite{10compute, 18complexity}. By taking advantage of quantum superposition, parallelism, and interference, inherent parallelism based on quantum state evolution can be achieved. This feature makes quantum computing particularly suitable for data-intensive machine learning tasks and spurs the emergence of quantum machine learning (QML) as an interdisciplinary field \cite{25network, 25network2, 17qml}. The main task of QML is to accelerate classical machine learning \cite{17qml}. For example, Grover's algorithm provides a quadratic speedup for unstructured database search \cite{96grover}, while quantum phase estimation (QPE) and quantum amplitude amplification (QAA) offer exponential acceleration potential for probability estimation and feature extraction \cite{02qae}. These capabilities enable QML to demonstrate advantages in tasks such as nearest neighbor search \cite{25shijieqknn}, kernel methods \cite{19kernel}, neural network \cite{26QADS}, and principal component analysis \cite{19pca}.

Inspired by these quantum advantages, in this work, we propose the quantum multi-label $k$-nearest neighbor (QML-$k$NN) algorithm, a novel framework that integrates quantum parallel computing with the local neighborhood similarity of ML-$k$NN to achieve significant acceleration while preserving and potentially enhancing classification performance. Specifically, QPE and controlled Grover iterations are first used to achieve parallel computation of prior probabilities, notably improving statistical efficiency. Then, an efficient quantum $k$-nearest neighbor search module is constructed by integrating the controlled-SWAP (c-SWAP) test with the quantum $k$-maximal search algorithm \cite{06durr, 18imageknn}, allowing rapid search of similar instances. Finally, we design a specialized quantum parallel counting circuit (QPCC) that operates by simultaneously collecting label frequencies from all training instances and their neighbors via quantum superposition. This enables the rapid extraction of posterior probabilities using quantum amplitude estimation (QAE) \cite{02qae}. The performance of the proposed QML-$k$NN is analyzed in terms of classification performance and time complexity, and its computational speedup is rigorously demonstrated. Experimental results show that QML-$k$NN significantly reduces time complexity while achieving superior prediction performance on established multi-label datasets. These findings indicate that QML-$k$NN provides an efficient solution for large-scale MLL tasks.

This paper is organized as follows. In Sec. \ref{Sec2}, we review the classical ML-$k$NN algorithm. In Sec. \ref{Sec3}, we present the design of the quantum circuit and the algorithmic workflow of the proposed QML-$k$NN framework. Performance analysis and discussion are presented in Sec. \ref{Sec4}, and conclusions are drawn in Sec. \ref{Sec5}.

\section{Multi-label $k$-nearest neighbor}
\label{Sec2}
ML-$k$NN is a lazy learning algorithm and a representative extension of traditional $k$NN to multi-label scenarios. As a lazy method, it foregoes explicit model training and directly exploits the entire training set to build a probabilistic model by counting label occurrences and co‑occurrences. For a new unseen instance, the algorithm then applies the maximum a posteriori (MAP) principle to its $k$ nearest neighbors to predict the proper set of labels.

Before setting up the probabilistic model, we first introduce some events. For a fixed label $l$, let $H_1^l$ denote the event that the instance has the label $l$ and $H_0^l$ the event that it does not. Furthermore, let $E_j^l$ (with $j\in\{0,1,...,k\}$) represent the event that, among the $k$ nearest neighbors of a given instance, exactly $j$ neighbors contain the label $l$.

The prior probability $P(H_1^l)$ is obtained from the training set by calculating the frequency $M_{l}$ of a label $l$. Using Laplace smoothing (with smoothing parameter $s$), we have
\begin{eqnarray}
\label{eq:prior}
P(H_1^l)&=&\dfrac{s+M_l}{s\times2+m}, \nonumber\\
P(H_0^l)&=&1-P(H_1^l).
\end{eqnarray}
where $m$ is the total number of instances.

The posterior probabilities refer to the probability of observing a specific neighbor‑label count given the instance’s own label status. For each label $l$ and each instance, the algorithm counts the number $j$ of its $k$ nearest neighbors that possess the label. This count is then routed based on whether the instance itself has the label $l$: if it does, the counter $c[j]$ is incremented and, otherwise, $c'[j]$ is incremented. After processing all instances, these accumulated counts directly give the required posterior probabilities
\begin{equation}\begin{array}{l}
\label{eq:posterior p}
P(E_{j}^{l}|H_{1}^{l}) = \dfrac{s + c[j]}{s(k + 1) + \sum_{p = 0}^{k} c[p]}= \dfrac{s + c[j]}{s(k + 1) + M_l}, \\
P(E_{j}^{l}|H_{0}^{l}) = \dfrac{s + c'[j]}{s(k + 1) + \sum_{p = 0}^{k} c'[p]}= \dfrac{s + c[j]}{s(k + 1) + m - M_l},
\end{array}\end{equation}
which completes the probabilistic model required for the prediction.

\begin{figure}[htbp]
\centering
\includegraphics[width=3.4in]{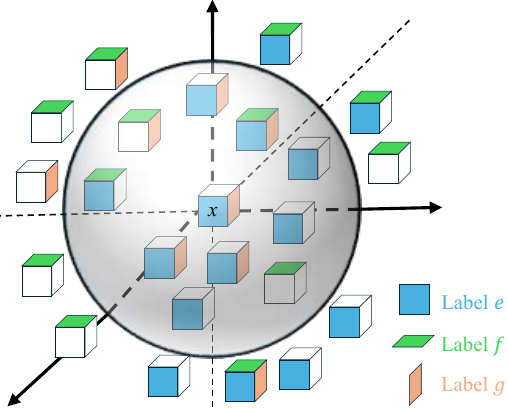}
\caption{An example of the ML-$k$NN algorithm. The dataset contains 23 instances, each possibly associated with up to three labels $e, f, g$ given in the legend. The circle marks the $k=10$ nearest neighbors of the central instance $x$.}
\label{shiyitu}
\end{figure}
A concrete example in Fig. \ref{shiyitu} is used to demonstrate how ML-$k$NN performs probability calculation, which includes 23 instances and three possible labels $e, f, g$. Among them, 15 instances have the label $e$, i.e., $M_{e}=15$, giving a prior probability $P(H_1^e)=(1+15)/(2+23)=0.64$, where $s=1$. Among $k=10$ nearest neighbors for a given instance $x$, 8 neighbors have the label $e$, and since this instance itself also carries this label, the count $c[8]$ increases by 1. Once such statistics have been collected for all instances, we still have $c[8] = 1$, and the posterior probability is calculated as $P(E_8^e|H_1^e)=(1+1)/[1 \times (10+1)+15]\approx 0.077$. Similarly, all posterior probabilities required for the probabilistic model can be obtained by their respective entries $c[j]$ and $c'[j]$.

Finally, for an unseen instance $t$, ML-$k$NN first identifies the $k$ nearest neighbors $N(t)$ of $t$ and counts, for each label $l$, the vector $\vec{C}_{t}(l)$ of neighbors containing $l$. Then the joint probabilities
\begin{equation}
\label{eq:pe}
P_b^l = P(H_b^l) \cdot P(E_{\vec{C}_{t}(l)}^{l}|H_{b}^{l}),
\end{equation}
are computed on the basis of this frequency vector. Using the Bayesian rule yields the following result
\begin{eqnarray}
\label{eq:bayesian}
P(H_1^l | E_{\vec{C}_{t}(l)}^{l}) &=& \frac{P_1^l}{P_1^l + P_0^l}, \nonumber\\
P(H_0^l | E_{\vec{C}_{t}(l)}^{l}) &=& \frac{P_0^l}{P_1^l + P_0^l}.
\end{eqnarray}
According to the MAP principle, the label $l$ is assigned to $t$ if $P(H_1^l | E_{\vec{C}_{t}(l)}^{l}) > P(H_0^l | E_{\vec{C}_{t}(l)}^{l})$. Repeating this for every label yields the final multi-label prediction.

\section{Quantum multi-label $k$-nearest neighbor}
\label{Sec3}
\begin{figure*}[htbp]
\centering
\includegraphics[width=1\textwidth]{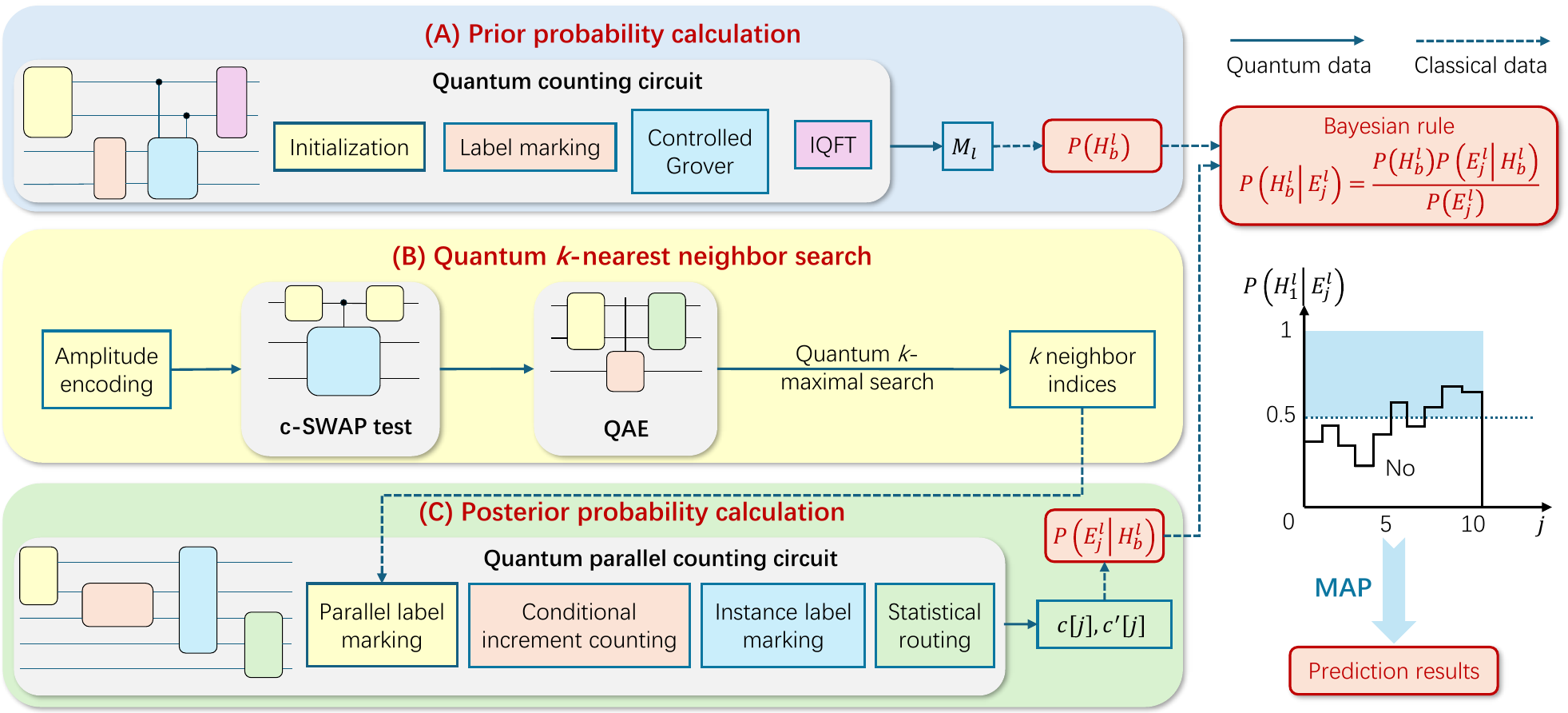}
\caption{Overall workflow of QML-$k$NN. It is divided into prior probability calculation, quantum $k$-nearest neighbor search and posterior probability calculation.}
\label{flowchart}
\end{figure*}

In this section, we detail the proposed QML-$k$NN, whose overall workflow is illustrated in Fig. \ref{flowchart}. It consists of three main modules, and each of them transforms the core component into a dedicated quantum subroutine that fundamentally differs from its classical counterpart. Module (A) employs a quantum counting circuit to efficiently calculate the prior probabilities for each label. Module (B) uses the c-SWAP test together with QAE and a quantum $k$-maximal search to rapidly identify the $k$ nearest neighbors of each given instance. Module (C) introduces a quantum parallel counting circuit (QPCC) that simultaneously collects label frequencies from all instances and their neighbors, enabling a fast calculation of the posterior probabilities. Each quantum subroutine of the proposed QML-$k$NN is detailed as follows.

\subsection{Prior probabilities calculation}
\label{Sec3.1}
\begin{figure}[h]
\centering
\includegraphics[width=3.5in]{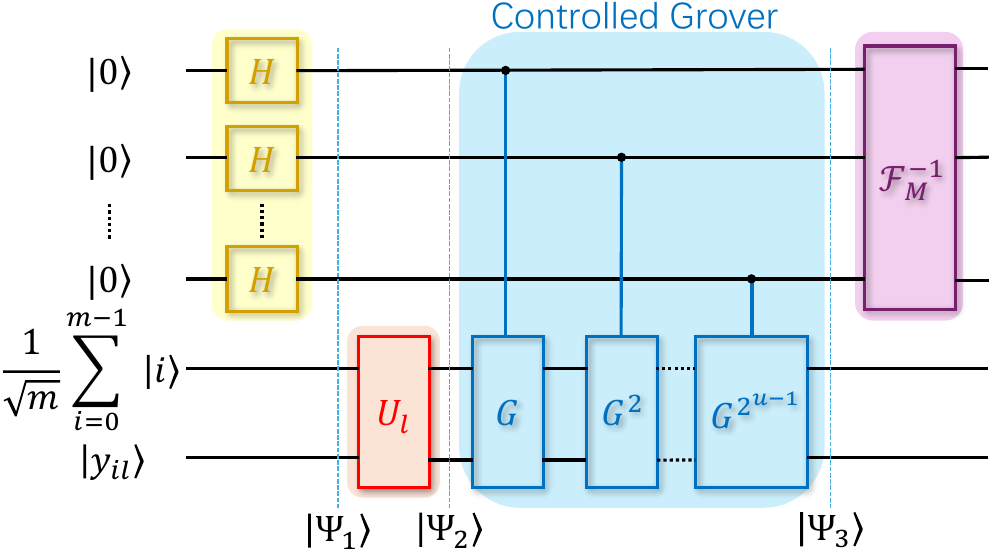}
\caption{Quantum counting circuit for calculating $M_{l}$. It mainly integrates controlled Grover iterations with IQFT, starting from a uniform superposition of all training instances.}
\label{prior}
\end{figure}

As shown in the quantum counting circuit of Fig. \ref{prior}, a quantum oracle $U_{l}$ first marks instances that carry the label $l$. The controlled Grover iterations then amplify the amplitudes of these marked instances, after which QPE with an inverse quantum Fourier transform (IQFT) $\mathcal{F}_M^{-1}$ extracts the encoded phase information. From the estimated phase, the frequency $M_{l}$ is obtained, and subsequently the prior probabilities are calculated.

In this circuit, specifically, the following quantum registers are first initialized: the instance index register $|i\rangle$ ($n=\lceil\log_{2}m\rceil$ qubits), the auxiliary label register $|y_{il}\rangle$ (a single qubit, where $y_{il}=1$ indicates that instance $i$ has label $l$ and $y_{il}=0$ otherwise), and the phase estimation register $|t\rangle$ ($u$ qubits). After applying Hadamard gates and CMP operation \cite{25shijieqknn}, the initial uniform superposition state is given by
\begin{equation}
|\Psi_1\rangle=\frac{1}{\sqrt{m2^u}}\sum_{i=0}^{m-1}\sum_{t=0}^{2^u-1}|i\rangle|0\rangle|t\rangle.
\end{equation}

A quantum oracle
\begin{equation}
U_{l} = \sum_{i=0}^{m-1} |i\rangle\langle i| \otimes X^{y_{il}},
\end{equation}
is then designed to query the classical label data and encode the value $y_{il}$ into the auxiliary qubit, yielding the quantum state
\begin{equation}
|\Psi_2\rangle=U_{l}|\Psi_1\rangle=
\frac{1}{\sqrt{m2^u}}\sum_{i,t}|i\rangle|y_{il}\rangle|t\rangle.
\end{equation}
It can also be expressed as \cite{97diffusion}
\begin{equation}
|\Psi_2\rangle=\sqrt{\frac{M_l}{m}}|\alpha\rangle|1\rangle+\sqrt{\frac{m-M_l}{m}}|\beta\rangle|0\rangle,
\end{equation}
where $|\alpha\rangle$ and $|\beta\rangle$ are superpositions of instances with and without label $l$, respectively. Let $\sin^2\theta = M_l / m$ $\left( \theta \in (0, \frac{\pi}{2}] \right)$, then the rotation angle of this state within the two-dimensional subspace is $\theta$. The Grover operator is constructed as 
\begin{equation} G=-U_l^\dagger S_0U_lS_\chi, \end{equation}
where $S_0$ obeys
\begin{equation}
\label{eq:S0}
S_0|x\rangle=
\begin{cases}
\phantom{-}|x\rangle, & x\neq 0 \hfill \\
-|x\rangle, & x=0 \hfill
\end{cases},
\end{equation}
and $S_{\chi}$ obeys
\begin{equation} S_{\chi}|i\rangle|y_{i,l}\rangle=(-1)^{y_{il}}|i\rangle|y_{il}\rangle. \end{equation}

The system then applies the powers of this operator $G^{2^t}$, controlled by the phase register $|t\rangle$. This process encodes the angle $\theta$ into the relative phase
\begin{align}
\lvert\Psi_{3}\rangle
  &= \frac{1}{\sqrt{m2^{u}}}
     \sum_{i,t}G^{2^{t}}\lvert i\rangle\lvert y_{il}\rangle\lvert t\rangle
     \nonumber\\
  &= \frac{1}{\sqrt{m2^{u}}}
     \sum_{t}\Bigl(
       e^{i2^{t+1}\theta}\sqrt{M_{l}}\lvert\alpha\rangle\lvert 1\rangle
     \nonumber\\
  &{}+ e^{-i2^{t+1}\theta}\sqrt{m-M_{l}}\lvert\beta\rangle\lvert 0\rangle
     \Bigr)\lvert t\rangle.
\end{align}

Subsequently, IQFT is applied to the register $|t\rangle$ to extract an estimate $\tilde{\phi} \approx 2\theta / \pi$. According to the principles of quantum counting, $M_l$ can be estimated as
\begin{equation}
\label{eq:Ml}
\widetilde{M_l}=m\cdot\sin^2{(\pi\tilde{\phi})}.
\end{equation}
The prior probabilities can be calculated using Eq. \eqref{eq:prior} and replace $M_l$ with $\widetilde{M_l}$.

\subsection{Quantum $k$-nearest neighbor search}
\label{Sec3.2}
With the prior probabilities obtained, the next step is to identify the $k$ nearest neighbors for each given instance, a prerequisite for the subsequent posterior probability calculation. To find the $k$ nearest neighbors of a given instance within the training set, pairwise similarity is calculated using the c‑SWAP test. This procedure encodes the fidelity between the given and each training instance in the amplitude of the ancilla qubit. The similarity values are then extracted through QAE and a quantum $k$-maximal search algorithm identifies the $k$ indices with the highest similarity.

In particular, the $U$-dimensional feature vectors of the given instance $x_i$ and of every training instance are first encoded into quantum states using amplitude encoding \cite{20data-encode}. The given $x_i$ state is prepared as \cite{14superposition-state}
\begin{equation}
|\rho\rangle = \frac{1}{\sqrt{U}}\sum_{i=1}^{U} |i\rangle\left( \sqrt{1 - v_{0i}^{2}}|0\rangle + v_{0i}|1\rangle \right)|1\rangle,
\end{equation}
and the superposition over the whole training set is encoded as
\begin{equation}
|\tau\rangle = \frac{1}{\sqrt{m}}\sum_{j=1}^{m} |j\rangle  \frac{1}{\sqrt{U}}\sum_{i=1}^{U} |i\rangle|1\rangle\left( \sqrt{1 - v_{ji}^{2}}|0\rangle + v_{ji}|1\rangle \right) ,
\end{equation}
where $v_{0i}$ and $v_{ji}$ denote the $i$-th feature component of the given instance and the $j$-th training instance.

\begin{figure}[h]
\centering
\includegraphics[width=1.8in]{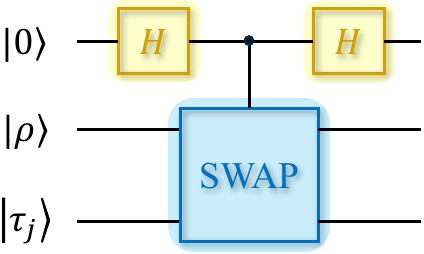}
\caption{Circuit for the c-SWAP test. It consists of two Hadamard gates and a SWAP gate that exchanges the two input states when the control qubit is $|1\rangle$.}
\label{swap}
\end{figure}

As shown in Fig. \ref{swap}, the c-SWAP test is then applied between $|\rho\rangle$ and each component $|\tau_j\rangle$ of $|\tau\rangle$. For the $j$-th training instance, the probability of measuring the control qubit in $|0\rangle$ is
\begin{equation}
P_j(0) = \frac{1 + |\langle\rho|\tau_j\rangle|^{2}}{2},
\end{equation}
which encodes the fidelity-based similarity $|\langle\rho|\tau_j\rangle|^{2}$. After this step, the overall state becomes
\begin{equation}
|\gamma\rangle = \frac{1}{\sqrt{m}}\sum_{j=1}^{m} |j\rangle \left( \sqrt{P_j(0)}|0\rangle + \sqrt{1 - P_j(0)}|1\rangle \right).
\end{equation}

\begin{figure}[h]
\centering
\includegraphics[width=2.3in]{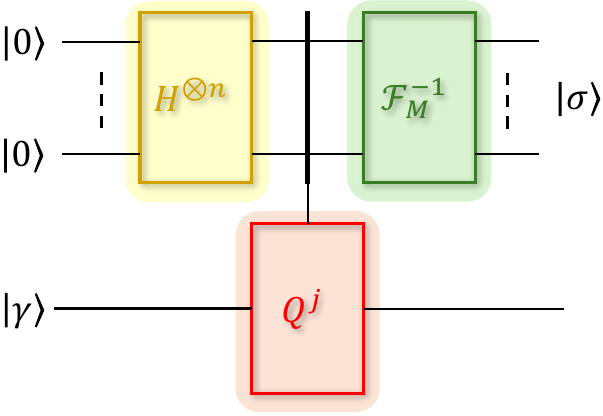}
\caption{The quantum circuit for QAE. QAA process converts the probability information associated with the amplitudes into corresponding phase information, which is subsequently extracted by the QPE procedure.}
\label{qae}	
\end{figure}

To obtain explicit similarity estimates, QAE is used to extract these probability values \cite{02qae}, which combines QAA (analogous to Grover iteration) with QPE, as illustrated in Fig. \ref{qae}. QAA is implemented by a unitary operator $Q = -\mathcal{A} S_0 \mathcal{A}^\dagger S_\chi$, where the operator $\mathcal{A}$ achieves the transformation $|0^{\otimes m}\rangle \rightarrow |\gamma\rangle$.

Upon applying the QAE module to the state $|\gamma\rangle$, we can obtain a new quantum state $|\sigma\rangle$ as
\begin{equation}
\begin{aligned}
|\sigma\rangle 
&= \frac{1}{\sqrt{m}}\sum_{j=1}^{m} |j\rangle\left| \frac{m}{\pi}\arcsin\left( \sqrt{\widetilde{P_j(0)}} \right) \right\rangle\\
&= \frac{1}{\sqrt{m}}\sum_{j=1}^{m} |j\rangle|\text{Sim}_j\rangle.
\end{aligned}\end{equation}

Consequently, each training instance index $|j\rangle$ is now associated not with a probability amplitude, but with a quantum register $|\text{Sim}_j\rangle$ that stores the estimated similarity value $\widetilde{P_j(0)}$. Since $\text{Sim}_j \propto |\langle\rho|\tau_j\rangle|^{2}$, it directly quantifies the similarity between instances. After that, the $k$ indices with the highest similarity are identified using the quantum $k$-maximal search algorithm, whose specific procedural steps are detailed in Algorithm~\ref{alg:qkm}. The resulting set $K$ of $k$ neighbor indices for the given instance $x_i$ is then used as input for the subsequent posterior probability calculation.

\begin{figure}[htbp]
\begin{algorithm}[H]
\caption{Quantum $k$-maximal search}
\label{alg:qkm}
\begin{algorithmic}[1]
\Require
  \Statex Quantum state $|\sigma\rangle$: superposition state containing $m$ similarity values;
  \Statex Integer $k$: number of maximum values to search for;
  \Statex Test instance $v_0$.
\Ensure
  \Statex Set $K$ containing $k$ indices of instances most similar to $v_0$.

\Function{QuantumMaxSearching}{$|\sigma\rangle, k, v_0$}
  \State Randomly select $k$ indices from $\{1,2,\dots,m\}$ as initial set $K$
  \State $K' \gets \{1,2,\dots,m\} \setminus K$
  \State $T \gets \lceil \sqrt{km}\, \rceil$  \Comment{Set max iterations}
  \For{$t = 1$ \textbf{to} $T$}
    \State Find index $i_{\max}$ in $K$ such that
    \State $d(v_0,v_{i_{\max}})=\max_{x\in\{1,\dots,k\}} d(v_0,v_{K_x})$
    \State Use Grover's algorithm to search for index $j\in K'$ satisfying
    \State $d(v_0,v_j) < d(v_0,v_{i_{\max}})$
    \If{such $j$ found}
      \State $K \gets (K \setminus \{i_{\max}\}) \cup \{j\}$
      \State $K' \gets (K' \setminus \{j\}) \cup \{i_{\max}\}$
    \Else
      \State \textbf{break}  \Comment{No better candidate}
    \EndIf
  \EndFor
  \State \Return $K$
\EndFunction
\end{algorithmic}
\end{algorithm}
\end{figure}

\subsection{Posterior probabilities calculation}
\label{Sec3.3}
To calculate posterior probabilities $P(E_j^l | H_b^l)$, a dedicated quantum parallel counting circuit (QPCC) is designed to calculate counters $c[j]$ and $c'[j]$ by performing label-frequency statistics in all training instances and their nearest neighbors simultaneously. 

\begin{figure*}[t]
\centering
\includegraphics[width=0.8\textwidth]{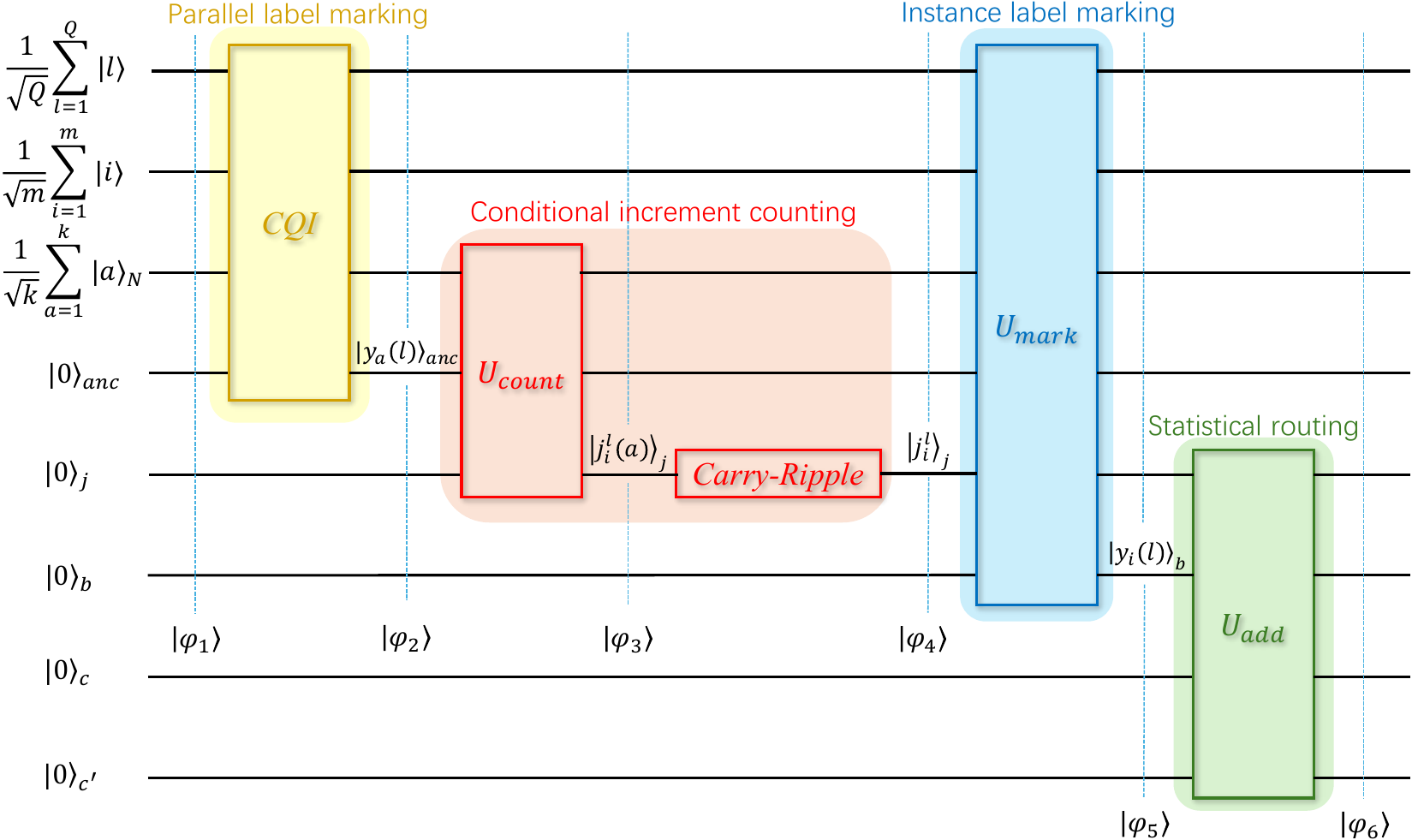}
\caption{Quantum parallel counting circuit for calculating $c[j]$ and $c'[j]$. The quantum registers used are: label index $|l\rangle$ ($\log Q$ qubits); instance index $|i\rangle$ ($\log m$ qubits); neighbor address $|a\rangle_{N}$ ($\log k$ qubits) for addressing the $k$ neighbors; auxiliary qubit $|0\rangle_{\mathrm{anc}}$ for marking the value of neighbor label; neighbor label count $|j\rangle_{j}$ ($\log (k+1)$ qubits) for counting how many neighbors possess the label; flag qubit $|0\rangle_{b}$ for marking whether the instance itself carries the label; and two statistics registers $|0\rangle_{c}$ and $|0\rangle_{c'}$ (each $\log m$ qubits) for tallying the labels occurrences of each count value among instances.}
\label{posterior}
\end{figure*}

As shown in Fig. \ref{posterior}, QPCC operates in four steps: parallel label marking, conditional increment counting, instance label marking, and statistical routing. The circuit uses a carry-ripple adder to accumulate counts conditionally and routes results based on whether the instance itself carries the label, subsequently estimating the required probability amplitudes.

To begin with, a uniform superposition state is prepared as
\begin{equation}
|\varphi_1\rangle = \frac{1}{\sqrt{Q m k}} \sum_{l=1}^{Q} \sum_{i=1}^{m} \sum_{a=1}^{k} |l\rangle |i\rangle |a\rangle_N |0\rangle_{anc} |0\rangle_j |0\rangle_b,
\end{equation}
which encompasses all labels, instances, and nearest neighbor indices. 

For each triple $(l, i, a)$, a classical-quantum interface (CQI) then marks whether the $a$-th neighbor of instance $i$ possesses the label $l$ and records the binary outcome $y_{a}(l)\in\{0,1\}$ in an auxiliary qubit, which produces the state
\begin{equation}
|\varphi_2\rangle = \frac{1}{\sqrt{Q m k}} \sum_{l,i,a} |l\rangle |i\rangle |a\rangle_N |y_a(l)\rangle_{anc} |0\rangle_j |0\rangle_b.
\end{equation}

A controlled-increment gate 
\begin{equation}
U_{count} = \bigoplus_{a=1}^{k} \left( |a\rangle\langle a|_N \otimes |1\rangle\langle 1|_{anc} \otimes \Lambda_j(+1) \right)
\end{equation}
is applied, which increments the counting register $|j\rangle_j$ only when $y_a(l) = 1$. After this global operation, the quantum state is transformed to
\begin{align}
|\varphi_3\rangle 
&=U_{count}|\varphi_2\rangle
\nonumber\\
&=\frac{1}{\sqrt{Q m k}} \sum_{l,i,a} |l\rangle |i\rangle |a\rangle_N |y_a(l)\rangle_{anc} |j_i^l(a)\rangle_j |0\rangle_b
\end{align}
where $|j_i^l(a)\rangle_j$ represents the cumulative count encoded in binary form after processing the first $a$ neighbors. The counting logic is implemented in superposition via a carry-ripple adder, detailed in the Appendix \ref{appendix:carry_ripple}. For each bit position $p$, a controlled carry gate is applied when all lower bits are 1 and the current neighbor is active, followed by a sum gate to update the current bit value. After processing all neighbors, the state becomes
\begin{equation}
\label{eq:phi4}
|\varphi_4\rangle = \frac{1}{\sqrt{Q m k}} \sum_{l,i,a} |l\rangle |i\rangle |a\rangle_N |y_a(l)\rangle_{anc} |j_i^l\rangle_j |0\rangle_b,
\end{equation}
where $j_i^l = \sum_{a=1}^{k} \delta(y_a(l), 1)$ is the total number of neighbors of instance $i$ that contain the label $l$.

After obtaining neighbor label counts, we mark whether instance $i$ itself possesses the label $l$ using a controlled-$X$ gate
\begin{equation}
U_{mark} = \sum_{l=1}^{Q} \sum_{i=1}^{m} |l\rangle\langle l| \otimes |i\rangle\langle i| \otimes I_j \otimes X^{y_i(l)} \otimes I_N \otimes I_{anc},
\end{equation}
which flips the flagging qubit $|0\rangle_b$ to $|1\rangle_b$ if $y_i(l) = 1$. Applying this gate to the state $|\varphi_4\rangle$, the resultant state can be expressed by
\begin{equation}
U_{mark}|\varphi_4\rangle = \frac{1}{\sqrt{Q m}} \sum_{i,l} |l\rangle |i\rangle |a\rangle_N |y_a(l)\rangle_{anc} |j_i^l\rangle_j |y_i(l)\rangle_b.
\end{equation}
Incorporating the initialized statistics registers, the system's quantum state can be expressed by
\begin{equation}
|\varphi_5\rangle = \frac{1}{\sqrt{m Q}} \sum_{l,i} |l\rangle |i\rangle |a\rangle_N |y_a(l)\rangle_{anc} |j_i^l\rangle_j |y_i(l)\rangle_b |0\rangle_c |0\rangle_{c'}.
\end{equation}

Subsequently, we design a controlled accumulation gate 
\begin{align}
U_{add} 
&= \sum_{j=0}^{k} \Bigl( |1\rangle\langle 1|_b \otimes |j\rangle\langle j|_j \otimes \Lambda_{c[j]}(+1) 
     \nonumber\\
&+ |0\rangle\langle 0|_b \otimes |j\rangle\langle j|_j \otimes \Lambda_{c'[j]}(+1) \Bigr)
\end{align}
to route the count $j_i^l$ to the appropriate statistics register based on $|b\rangle$. If $b = 1$, indicating that the instance has the label $l$, the count accumulates into $|c[j]\rangle_c$; otherwise, into $|c'[j]\rangle_{c'}$. The final quantum state contains complete statistical information, which can be expressed by
\begin{align}
|\varphi_6\rangle
&=\frac{1}{\sqrt{mQ}}\sum_{l,i}|l\rangle|i\rangle|a\rangle_N|y_a(l)\rangle_{anc}|j_i^l\rangle_j|y_i(l)\rangle_b
     \nonumber\\
&\otimes|c[j_i^l]+\delta_{b,1}\rangle_c|c^{\prime}[j_i^l]+\delta_{b,0}\rangle_{c^{\prime}}.
\end{align}

For each $j$, we use QAE to concurrently estimate the probability amplitudes for all $(l, i)$ pairs, obtaining the temporary probability value
\begin{equation}
\widehat{P}(c[j] = n) = \left| \frac{1}{\sqrt{mQ}}\sum_{(l,i) \in S_{j,n}} e^{i\theta_{l,i}} \right|^{2},
\end{equation}
where $S_{j,n} = \{(l,i) \mid j_i^{l} = j \land y_i(l) = 1\}$ defines the set of valid instance-label pairs and $\theta_{i,l}$ represents a phase rotation parameter. When the number of QAE iterations $M$ satisfies $M \geq \pi\sqrt{mQ/\epsilon}$, the estimation error $\epsilon$ becomes negligible. Ensuring coverage across all values of $j$, the counts are then derived as $c[n] = mQ \cdot \widehat{P}(c[j] = n)$. From Eqs. \eqref{eq:posterior p} and \eqref{eq:Ml}, the posterior probabilities $P(E_{j}^{l}|H_{b}^{l})$ are calculated based on all the counting results $c[j]$ and $c'[j]$.

After obtaining the probabilistic model, predicting an unseen instance $t$ follows the same MAP principle using the Bayesian rule in Eq. \eqref{eq:bayesian}.

\section{Performance analysis and discussion}
\label{Sec4}
Having detailed the framework of QML-$k$NN, we now analyze its performance in terms of classification performance and time complexity.

\subsection{Classification Performance}
\label{Sec4.1}
\begin{figure}[h]
\centering
\includegraphics[width=3.4in]{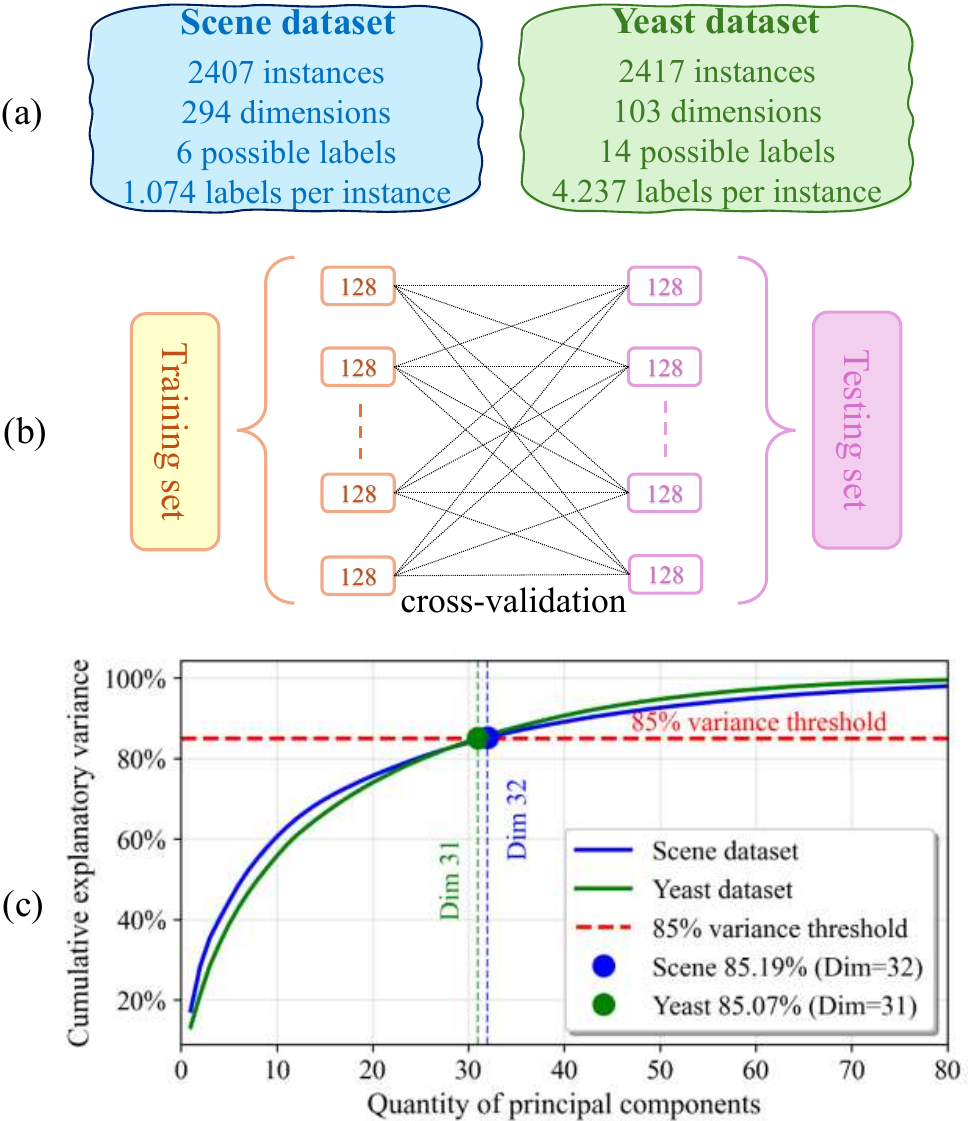}
\caption{Overview of the experimental setup on the dataset. (a) Dataset characteristics; (b) Random sampling; (c) PCA for feature dimensionality reduction.}
\label{dataset}	
\end{figure}

Here, we first evaluate the performance of QML-$k$NN on two established multi-label datasets: Scene (natural image classification) and Yeast (gene functions of Saccharomyces cerevisiae) \cite{04pattern, 01gene-kernel}. Their key characteristics, including the number of instances, feature dimensions, labels, and the cardinality (average number of labels per instance), are summarized in Fig. \ref{dataset}(a). Scene contains 2407 instances with 294‑dimensional features and 6 labels, having a low average cardinality of 1.074. Yeast comprises 2417 instances with 103‑dimensional features and 14 labels, exhibiting a much higher average cardinality of 4.237 and consequently denser label co‑occurrence patterns.

Before conducting experiments, we apply two preprocessing steps to both datasets to ensure robust evaluation and computational feasibility. First, to manage the large number of instances and enable robust evaluation, we randomly extract subsamples from the two pre-partitioned sets of each dataset and apply the cross-validation strategy, as illustrated in Fig. \ref{dataset}(b). For Scene, nine subsampling rounds are implemented: 128 instances are randomly selected from the original training set (1211 instances), and another 128 from the test set (1196 instances) in each round. For Yeast, seven rounds are performed, with 128 instances subsampled per round from its training set (1500 instances) and test set (917 instances). Second, to address the high dimensionality of the features, we apply principal component analysis (PCA) to reduce the dimension for computational feasibility \cite{10pca}. As shown in Fig. \ref{dataset}(c), we reduce the feature dimension to 32 for the Scene and to 31 for the Yeast, both retaining more than 85\% of the original explained variance. This allows efficient model training without substantial information loss for both datasets.

We first evaluate the algorithms in the Scene dataset. For each of the nine random subsampling rounds, we train the probabilistic models in the 128 training instances and test in the corresponding 128 test instances. The mean and standard deviation of six metrics across the nine rounds are listed in Table. \ref{scene}.
\begin{table}[htbp]
\centering
\caption{Experimental Results Based on Random Subsampling of the Scene Dataset (mean ± std).}
\label{scene}
\begin{tabular}{lcc}
\hline
\multirow{2}{*}{Evaluation metrics} & \multicolumn{2}{c}{Algorithm} \\
\cline{2-3}
 & QML-$k$NN & ML-$k$NN\\
\hline
Hamming Loss & 0.141±0.016 & \textbf{0.140±0.010}\\
One-error & \textbf{0.343±0.102} & 0.364±0.046\\
Coverage & 0.992±0.229 & \textbf{0.786±0.087}\\
Ranking Loss & 0.144±0.050 & \textbf{0.137±0.020}\\
Average Precision & \textbf{0.784±0.073} & 0.778±0.027\\
Macro-AUC & 0.813±0.027 & \textbf{0.861±0.022}\\
\hline
\end{tabular}
\end{table}

Specifically, QML-$k$NN achieves lower One-error and higher Average Precision than classical ML-$k$NN. Both metrics are top‑weighted, and they mainly evaluate whether the most relevant labels are placed at the top of the ranking, without requiring the exact classification of every label \cite{07ml-knn}. QML-$k$NN excels in this regard because its quantum parallelism enables the simultaneous collection of statistics across all instances, labels, and neighbors. This global processing better preserves the relative order among labels by capturing co‑occurrence patterns that sequential counting may miss, leading to more accurate top‑rank decisions even when absolute probability estimates contain minor errors.

For the remaining metrics, QML-$k$NN performs slightly worse than classical ML-$k$NN on Coverage, Ranking Loss, and Macro-AUC, with comparable Hamming Loss. Hamming Loss and Macro-AUC are threshold-sensitive: the former depends on per-label thresholding, while the latter averages binary AUCs across all labels, making both susceptible to small probability perturbations. Coverage requires the exact ranking position of the last relevant label, which is vulnerable to ordering noise. The QAE in QML-$k$NN introduces small probability perturbations that, while harmless for top‑order decisions, can shift a label across a threshold or alter its rank position when all labels must be considered. The exact frequency counting of ML-$k$NN provides deterministic estimates that are more stable for these strict requirements. This stability gives ML-$k$NN its advantage in exact multi‑label classification matches, where every label's prediction must be correct.

Based on the above analysis, the main advantage of QML-$k$NN lies in its ability to place the most relevant labels at the top of the ranking, while ML‑$k$NN retains an advantage in exact multi-label classification matches. It is worth noting that the Scene dataset has a low cardinality of only 1.074, meaning that most scene images are associated with a single label. In such a setting, the multi-label nature of the problem is not fully exercised, and the potential of QML-$k$NN in handling complex label interactions may not be adequately revealed.

To gain a deeper evaluation of the proposed algorithm, we further conduct experiments on the Yeast dataset, which is characterized by a higher cardinality of 4.237, following the same evaluation method. As shown in Table \ref{yeast7groups}, it can be easily observed that QML-$k$NN outperforms ML-$k$NN in terms of all metrics, which distinctly differs from the results in Scene.

\begin{table}[htbp]
\centering
\caption{Experimental Results Based on Random Subsampling of the Yeast Dataset (mean ± std).}
\label{yeast7groups}
\begin{tabular}{lcc}
\hline
\multirow{2}{*}{Evaluation metrics} & \multicolumn{2}{c}{Algorithm} \\
\cline{2-3}
 & QML-$k$NN & ML-$k$NN\\
\hline
Hamming Loss & \textbf{0.219±0.027} & 0.229±0.012\\
One-error & \textbf{0.250±0.033} & 0.265±0.042\\
Coverage & \textbf{6.674±0.294} & 7.085±0.192\\
Ranking Loss & \textbf{0.185±0.013} & 0.211±0.021\\
Average Precision & \textbf{0.727±0.013} & 0.714±0.025\\
Macro-AUC & \textbf{0.609±0.041} & 0.557±0.037\\
\hline
\end{tabular}
\end{table}

Specifically, for the four ranking metrics, i.e., One-error, Coverage, Ranking Loss, and Average Precision, QML-$k$NN achieves consistent improvements, indicating a more accurate relative ordering among labels. For the two classification metrics, Hamming Loss and Macro‑AUC, QML-$k$NN also shows clear gains. The superior Macro‑AUC further confirms its improved binary discriminative ability across all labels, including the less frequent ones.

\begin{figure*}[htbp]
\centering
\includegraphics[width=1\textwidth]{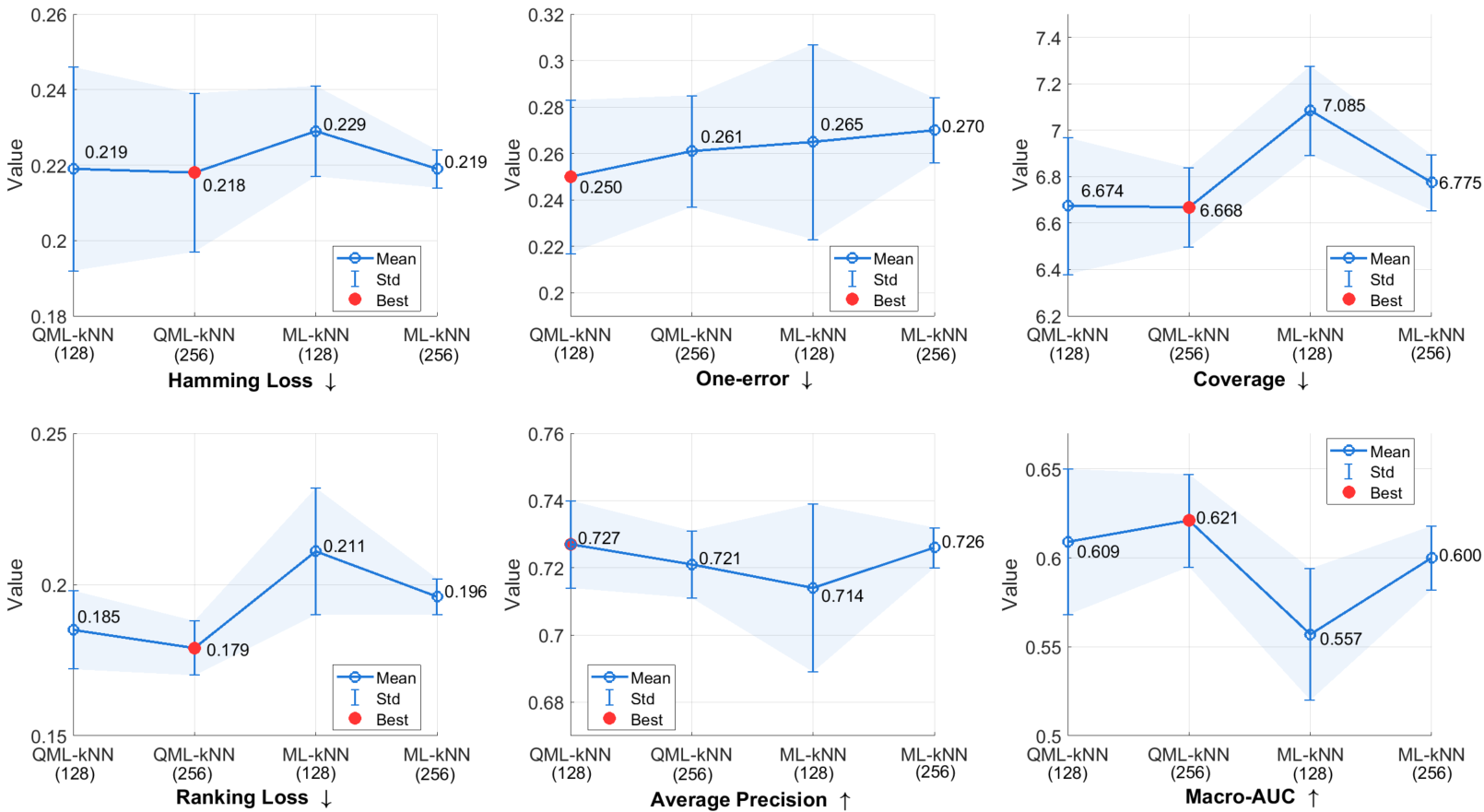}
\caption{Experimental results on Yeast dataset. Lines represent mean values of each evaluation metric, shaded areas indicate standard deviations, and red markers highlight the best value among the three algorithms.}
\label{evaluation}
\end{figure*}

This comprehensive superiority can be attributed to a fundamental difference in how the two algorithms search nearest neighbors and collect label statistics. In classical ML-$k$NN, the $k$ nearest neighbors of each instance are identified by computing Euclidean distances sequentially, a process that becomes increasingly unreliable in high‑dimensional feature spaces due to the well‑known curse of dimensionality. Prior studies have shown that quantum‑enhanced $k$NN algorithms can achieve significant improvements in classification accuracy, especially for high‑dimensional datasets, as quantum similarity measures based on fidelity can more effectively capture the underlying structure of such spaces \cite{14superposition-state, 21qknn}. QML-$k$NN leverages this advantage through two key mechanisms. First, the c‑SWAP test computes similarities based on fidelity between quantum‑encoded feature vectors, providing a more robust similarity assessment in the high dimensional biological feature space of the Yeast. Second, the QPCC performs label statistics collection in parallel across all labels, instances, and neighbors within a single coherent superposition, rather than decomposing the problem into independent per‑label estimations. This parallel processing ensures that the correlations among labels, i.e., the tendency of certain functional classes to co‑occur, are naturally preserved throughout the counting process. As a result, the posterior probabilities $P(E_j^l \mid H_b^l)$ derived from these globally collected statistics better reflect the underlying correlation structure among labels, leading to consistent improvements across all metrics.

Moreover, the quantum advantage is obvious on Yeast because the high label density provides rich correlation structures for the quantum parallelism. The more complex label interactions, the greater the benefit of moving from independent estimation per‑label to global informed probability modeling. This is in contrast to the Scene, where low label density offered limited correlation structure and the quantum advantage in ranking did not fully compensate for the estimation noise.

Given that the above results use only 128 training instances, we are wondering that can ML-$k$NN make up for this performance gap with more data? To investigate this, we conduct an additional experiment in which the number of training and testing instances per round is increased from 128 to 256, while keeping all other settings unchanged. The results for QML-$k$NN (256) and ML-$k$NN (256) are shown in Fig. \ref{evaluation}. As expected, increasing the sample size improves most evaluation metrics for ML-$k$NN. For instance, Hamming Loss drops from 0.229 to 0.219, Coverage from 7.085 to 6.775, and Ranking Loss from 0.211 to 0.196. The only metric where ML-$k$NN (256) achieves a slightly worse value than ML-$k$NN (128) is the One-error (0.270 vs. 0.265), and this is attributed to overfitting due to class imbalance. 

For QML-$k$NN, expanding the training set from 128 to 256 also yields some improvements, particularly in Ranking loss (from 0.185 to 0.179) and Macro‑AUC (from 0.609 to 0.621). However, the gains are relatively modest compared to the doubling of training data. In particular, for metrics such as Hamming Loss (from 0.219 to 0.218), One-error (from 0.250 to 0.261) and Average Precision (from 0.727 to 0.721), the improvements are less and even worse, suggesting that QML-$k$NN already achieves strong performance with a small training set.

More importantly, even with twice the amount of training data, ML-$k$NN (256) still fails to surpass QML-$k$NN (128) in almost all metrics. In terms of Hamming Loss, the performance of the two algorithms is nearly identical. In all other five metrics, QML-$k$NN (128) remains superior to ML-$k$NN (256), with particularly notable advantages in One‑error (0.250 vs. 0.270) and Ranking Loss (0.185 vs. 0.196).

These findings indicate that the performance advantage of QML-$k$NN is not simply due to an insufficient sample size for ML-$k$NN. Instead, the quantum parallelism inherent in QML-$k$NN's probability calculation and neighbor search allows it to extract more informative patterns from limited data, achieving better predictive accuracy even with fewer training instances. In other words, QML-$k$NN demonstrates superior data efficiency compared to ML-$k$NN. Furthermore, QML-$k$NN shows only minor improvements when training data are doubled, which further confirms that the quantum mechanism effectively captures the essential statistical structure of the data, and that additional training examples provide fewer returns once the quantum similarity measure has identified the most informative patterns.

This data efficiency is particularly valuable on the Yeast dataset, which exhibits high label density and complex nonlinear feature relationships. The features, derived from micro-array expression and phylogenetic profiles, are inherently noisier and embody more complex biological relationships than hand-made feature sets \cite{01gene-kernel}. Quantum parallelism in QML-$k$NN provides a more effective mechanism for learning these high-order dependencies. Furthermore, the evolution of quantum states and interference mechanisms contribute to a more robust feature representation and a more effective assessment of similarity in this challenging and high-dimensional biological feature space \cite{19kernel}. This leads to improved nearest neighbor search and, consequently, to a more accurate probabilistic model.

In summary, these experimental results demonstrate that the advantage of QML-$k$NN becomes particularly suitable in datasets characterized by high label density and complex, nonlinear feature-label relationships. Its superior data efficiency and more effective similarity assessment yield comprehensive performance gains even with limited training data.

\subsection{Time Complexity}
\label{Sec4.2}
Below we investigate the time complexity of the proposed QML-$k$NN and compare it with the classical ML-$k$NN. Table. \ref{tab:posterior time_complexity} summarizes the leading-order time complexity of each algorithmic step for both algorithms, and the detailed derivations are as follows.
\begin{table*}[htbp]
\centering
\caption{Time Complexity of Individual Steps in Posterior Probability Computation}
\label{tab:posterior time_complexity}
\begin{tabular}{lcc}
\hline
Operation & ML-$k$NN & QML-$k$NN \\
\hline
Prior probability calculation & $O(QM)$ & $O(Q\sqrt{m})$ \\
Similarity calculation & $O(m^2U)$ & $O(m^2\log^2U)$ \\
$k$-nearest neighbor search & $O(m^2\log m)$ & $O(m^{1.5}\sqrt{k})$ \\
Posterior probability calculation & $O(Qk(m+1))$ & $O(m\log k + (Q + m )/\epsilon)$\\
\hline
\end{tabular}
\end{table*}

In particular, we first analyze the prior probability calculation. In ML-$k$NN, computing the prior probability for each label requires scanning all $m$ instances to count the occurrences, resulting in a time complexity of $O(Qm)$. However, QML-$k$NN uses the quantum counting circuit to estimate the frequency for each label. Starting from the state $|\Psi_2\rangle = \sqrt{M_l/m} |\alpha\rangle|1\rangle + \sqrt{(m-M_l)/m} |\beta\rangle|0\rangle$, a single Grover iteration $G$ rotates the state vector by an angle $2\theta$ towards the superposition of the solution states $|\alpha\rangle$, where $\sin\theta = \sqrt{M_l/m}$. The IQFT then extracts the phase, requiring $R$ rotations that satisfy
\begin{equation}
R \cdot 2\theta \approx \pi/2 \implies R \approx \frac{\pi}{4\theta} \approx \frac{\pi}{4} \sqrt{\frac{m}{M_l}}.
\end{equation}
Therefore, the time complexity per label is $O\left( \sqrt{m/M_l} \right)$, resulting in an overall complexity of $O\left(Q \cdot \sqrt{m/min(M_l)}\right)$. In the worst-case scenario where $\min(M_l) = 1$, the total complexity becomes $O(Q\sqrt{m})$. Under average conditions, where $\min(M_l) = \Theta(1)$, the overall complexity remains $O(Q\sqrt{m})$.

We then analyze the similarity calculation for the nearest neighbor search. In ML-$k$NN, computing the similarity between a given instance and all other training instances costs $O(mU)$. Repeating this for all $m$ instances leads to $O(m^2 U)$. In contrast, QML-$k$NN executes a single c‑SWAP test with complexity $O(\log^2 U)$ \cite{22cswap}. Performing $m$ such tests in parallel for each training instance yields $O(m^2 \log^2 U)$. Estimating similarity amplitudes via QAE requires $O(1/\epsilon)$ repetitions of the Grover operator. Given that each Grover operator has a complexity of $O(1)$, the overall complexity of QAE is $O(m / \epsilon)$. Thus, the dominant complexity for this part mainly amounts to $O(m^2 \log^2 U)$, achieving a logarithmic dependence on feature dimensionality compared to classical linear scaling.

Next, we consider the $k$ nearest neighbors search from the calculated similarity values. In ML-$k$NN, sorting the $m$ similarity values for each instance (e.g., using quicksort or heapsort) adds $O(m^2 \log m)$ to the overall cost  \cite{94sorting}. In QML-$k$NN, the quantum $k$-maximal search algorithm identifies the $k$ largest similarities among the $m$ instances with complexity $O(\sqrt{km})$ \cite{06durr}, leading to an overall complexity of $O(m^{1.5} \sqrt{k})$ for this identification stage.

In terms of the posterior probability calculation, ML-$k$NN iterates through every training instance, for each label $l$, counting how many of its $k$ nearest neighbors possess the label, resulting in a complexity of $O(Qmk)$. Subsequently, the normalization of posterior probabilities, i.e., the calculation of $P(E_j^l |H_1^l)$ and $P(E_j^l |H_0^l)$ for all labels incurs a cost of $O(Qk)$. Consequently, the aggregate complexity becomes $O(Qk(m+1))$. By contrast, QML-$k$NN performs the posterior probability calculation via QPCC as follows. The loading of $k$ neighbor indices through the CQI takes $O(1)$ time. Conditional increment counting, implemented using a carry-ripple adder, has a complexity of $O(\log k)$. Instance label marking and statistical routing mainly involve controlled operations that incur $O(1/ \epsilon)$ time. Taking these into account, the statistical time per single instance is $O(\log k + 1/ \epsilon)$, leading to a total time of $O(m(\log k + 1/ \epsilon))$ for all instances. Subsequently, the normalization of posterior probabilities is performed on the accumulated counts with a complexity of $O(Q/ \epsilon)$, leading to an overall complexity of $O(m\log k + (Q + m )/\epsilon)$.

\begin{figure}[h]
\centering
\includegraphics[width=3.4in]{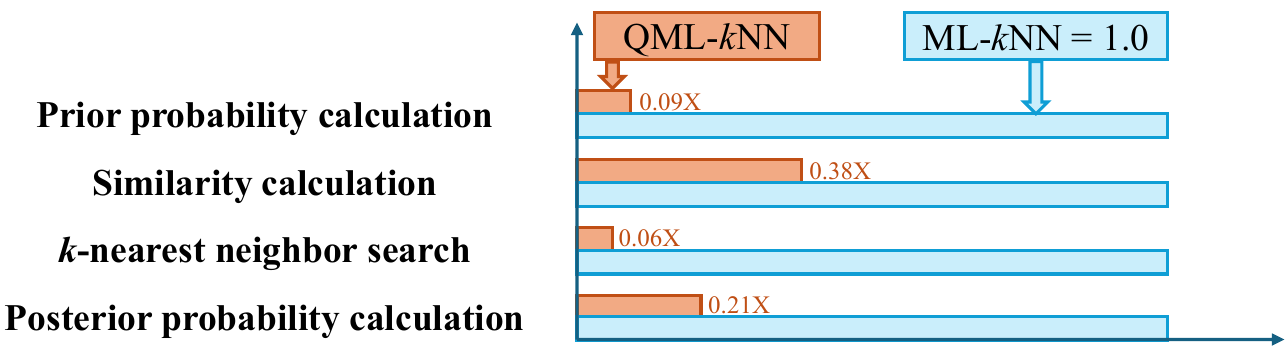}
\caption{The leading-order complexity terms. Based on the Yeast experimental setup, parameters are set to be as follows. $U = 31$, $m = 128$, $Q = 14$, $k = 10$ and $\epsilon = 0.1$. Each step is normalized separately, i.e., ML-$k$NN = 1.0.}
\label{radar}
\end{figure}

To visualize the performance gap, we take the typical parameter values from our experiments into the complexity expressions. The resulting leading‑order terms are plotted in Fig. \ref{radar}, where each step is normalized separately with ML-$k$NN set to 1.0. It clearly shows that QML-$k$NN achieves an obvious reduction in all steps.

\begin{figure*}[htbp]
\centering
\includegraphics[width=0.7\textwidth]{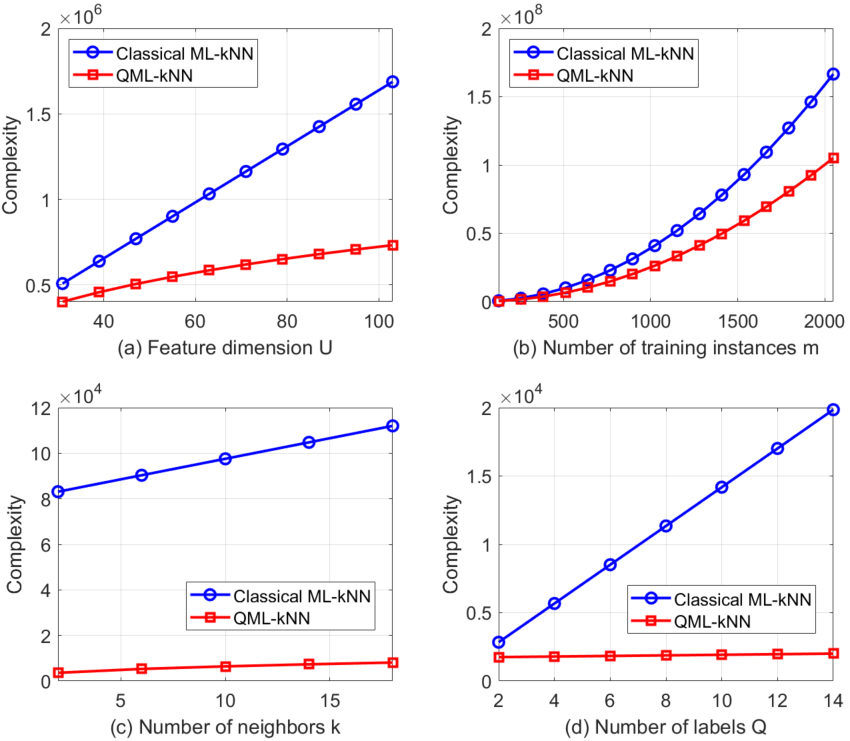}
\caption{Complexity comparisons as a function of dimension $U$ (a), training instances $m$ (b), neighbors $k$ (c), and labels $Q$ (d).}
\label{parameter}
\end{figure*}
Furthermore, to systematically characterize the scaling behavior of the time complexity advantage of QML-$k$NN over classical ML-$k$NN, we conduct a controlled parametric analysis in which each variable is varied independently while the others are kept constant. Fig. \ref{parameter}(a) shows that the feature dimension $U$ exhibits the most fundamental difference in scaling. ML-$k$NN incurs a linear complexity cost $O(m^2 U)$ for pairwise similarity calculation, while the c-SWAP test in QML-$k$NN leverages the amplitude encoding to achieve logarithmic dependence on $U$, resulting in a rapidly widening complexity gap as the dimension increases. As shown in Fig. \ref{parameter}(b), the size of the training set $m$ represents the dominant computational bottleneck for both algorithms, with ML-$k$NN showing superlinear growth $O(m^2 U+m^2 \log m)$ due to exhaustive neighbor search and sorting. In contrast, QML-$k$NN reduces this scaling to $O(m^2 \log^2 U + m^{1.5} \sqrt{k})$ through the quantum $k$-maximal search and achieves a quadratic speedup $O(Q\sqrt{m})$ for prior probability estimation via quantum counting, leading to a dramatic expansion of the performance gap. Fig. \ref{parameter}(c) shows that the hyper‑parameter $k$ has minimal impact on the relative complexity advantage of QML-$k$NN. Since majority voting constitutes only a small fraction of the total cost for both algorithms, resulting in stable acceleration in all the values tested for $k$. As shown in Fig. \ref{parameter}(d), for the size $Q$ of the set of labels, ML-$k$NN suffers from a linear scaling $O(Qk(m+1))$ in the posterior probability calculation. However, in QML-$k$NN, QPCC eliminates the linear dependence on $m$ and reduces the cost to $O(m\log k + ( m + Q ) / \epsilon)$. 

In general, these results demonstrate that the theoretical speedups of QML-$k$NN are consistent and become significant as the scale of dataset increases. Although practical speedup is constrained by current quantum hardware limitations, this theoretical advantage provides the substantial potential of QML-$k$NN to efficiently process large-scale datasets.

\section{Conclusion}
\label{Sec5}
In this work, we have proposed a multi-label $k$-nearest neighbor algorithm, called QML-$k$NN. By reformulating the core steps of ML-$k$NN into a quantum framework, the proposed algorithm leverages fundamental quantum properties, including superposition, parallelism, and interference, to address the computational efficiency bottlenecks inherent in traditional approaches when dealing with large-scale multi-label data.

Theoretically, QML-$k$NN accelerates the calculation of prior probabilities through a quantum counting circuit, thus reducing the time complexity. Furthermore, the algorithm uses the c-SWAP test to calculate instance similarity and adopts a quantum $k$-maximal search algorithm to efficiently identify nearest neighbors, achieving significant acceleration in this module. Finally, the designed QPCC enables fast computation of posterior probabilities.

The experimental results reveal that QML-$k$NN achieves comprehensive performance superiority. It outperforms ML-$k$NN across all evaluation metrics, including Hamming Loss, ranking-based measures, and Macro-AUC. Notably, even when the training set is doubled, ML-$k$NN still cannot surpass QML-$k$NN trained on only half of the data, and expanding QML-$k$NN's own training set yields only marginal improvements, indicating that additional training examples provide fewer returns. This notable performance gain, combined with the theoretical reduction in time complexity, demonstrates the inherent advantages of the quantum paradigm. The results indicate that the quantum advantage of QML-$k$NN becomes especially prominent on datasets with intricate label interactions and high-dimensional noisy feature spaces, as it captures and exploits high-order label dependencies more effectively via quantum parallelism.

This study verifies the great potential of quantum computing for complex machine learning tasks, especially multi-label learning. Future work will focus on deploying the proposed algorithmic modules on near-term and mid-term quantum hardware, as well as further optimizing the framework to cut quantum resource overhead.

\appendix
\section{Carry-Ripple Adder}
\label{appendix:carry_ripple}
For the conditional increment counting operation inside QPCC, a carry-ripple adder is adopted to realize efficient binary counting in the quantum domain.

First, the neighbor count register $|j\rangle_j$ is decomposed into $n = \lceil \log(k + 1) \rceil$ qubits, denoted as $|j\rangle_j = |j_{n-1}\rangle \otimes \cdots \otimes |j_0\rangle$, which represents the binary integer $j = \sum_{p=0}^{n-1} 2^p j_p$. For each neighbor $a$ and each bit position $p$, a controlled carry gate
\begin{equation}
\text{Carry}_p = |a\rangle\langle a|_N \otimes |1\rangle\langle 1|_{anc} \otimes \left( \bigotimes_{q=0}^{p-1} |1\rangle\langle 1|_{j_q} \right) \otimes X_{j_p}.
\end{equation}
is applied. It flips the qubit $j_p$ to simulate a carry operation if and only if all the following conditions hold simultaneously: the currently processed neighbor is indexed by $a$, this neighbor carries label $l$ (i.e., $y_a(l) = 1$), and all low-order bits $j_0, \ldots, j_{p-1}$ stay in state $|1\rangle$.

Subsequently, a sum gate is introduced to update each target bit:
\begin{equation}
\text{Sum}_p = |a\rangle\langle a|_N \otimes |1\rangle\langle 1|_{anc} \otimes X_{j_p},
\end{equation}
which directly flips $j_p$ under the condition $y_a(l) = 1$ and completes the bit update for accumulation.

Since the quantum superposition covers all neighbor indices $a$, the composite counting unitary operates in parallel over all $a$:
\begin{equation}
U_{CS} = \prod_{p=0}^{n-1} \left( \prod_{a=1}^{k} \text{Carry}_p^{(a)} \right) \cdot \left( \prod_{a=1}^{k} \text{Sum}_p^{(a)} \right).
\end{equation}
After implementing unitary $U_{CS}$, the evolved quantum state equals $|\varphi_4\rangle$ (Eq.~\eqref{eq:phi4} in the main text).

In summary, the carry-ripple adder utilizes quantum superposition to execute conditional increment counting for all instances and labels in parallel, generating the basic statistical data required for the posterior probability routing module. This unitary circuit realizes efficient and logically consistent binary counting directly within quantum states.

\section*{Acknowledgments}
This work was supported by the Hunan Provincial Natural Science Foundation of China (Grant No. 2026JJ40054), the Hunan Provincial Key Research and Development Program (Grant No. 2025QK3011) and the National Natural Science Foundation of China (Grant No. 62101180).

\bibliographystyle{apsrev4-1}
\bibliography{references}

@inproceedings{07supervised,
  title={Supervised Machine Learning: A Review of Classification Techniques},
  author={Sotiris B. Kotsiantis},
  booktitle={Informatica},
  year={2007},
  url={https://api.semanticscholar.org/CorpusID:47128183}
}

@ARTICLE{14mll-review,
  author={Zhang, Min-Ling and Zhou, Zhi-Hua},
  journal={IEEE Transactions on Knowledge and Data Engineering}, 
  title={A Review on Multi-Label Learning Algorithms}, 
  year={2014},
  volume={26},
  number={8},
  pages={1819-1837},
  doi={10.1109/TKDE.2013.39}}

@ARTICLE{22mll,
  author={Liu, Weiwei and Wang, Haobo and Shen, Xiaobo and Tsang, Ivor W.},
  journal={IEEE Transactions on Pattern Analysis and Machine Intelligence}, 
  title={The Emerging Trends of Multi-Label Learning}, 
  year={2022},
  volume={44},
  number={11},
  pages={7955-7974},
  doi={10.1109/TPAMI.2021.3119334}}

@article{04pattern,
title = {Learning multi-label scene classification},
journal = {Pattern Recognition},
volume = {37},
number = {9},
pages = {1757-1771},
year = {2004},
issn = {0031-3203},
doi = {https://doi.org/10.1016/j.patcog.2004.03.009},
url = {https://www.sciencedirect.com/science/article/pii/S0031320304001074},
author = {Matthew R. Boutell and Jiebo Luo and Xipeng Shen and Christopher M. Brown}
}

@article{07ml-knn,
title = {ML-KNN: A lazy learning approach to multi-label learning},
journal = {Pattern Recognition},
volume = {40},
number = {7},
pages = {2038-2048},
year = {2007},
issn = {0031-3203},
doi = {https://doi.org/10.1016/j.patcog.2006.12.019},
url = {https://www.sciencedirect.com/science/article/pii/S0031320307000027},
author = {Min-Ling Zhang and Zhi-Hua Zhou}
}

@article{25network,
title = {Quantum generative adversarial network based on the quantum Born machine},
journal = {Advanced Engineering Informatics},
volume = {68},
pages = {103622},
year = {2025},
issn = {1474-0346},
doi = {https://doi.org/10.1016/j.aei.2025.103622},
url = {https://www.sciencedirect.com/science/article/pii/S1474034625005154},
author = {Yifan Ding and Zi Li and Nanrun Zhou}
}

@article{25network2,
author = {Gong, Li-Hua and Chen, Yu-Qi and Zhou, Shun and Zeng, Qing-Wei},
title = {Dual Discriminators Quantum Generation Adversarial Network Based on Quantum Convolutional Neural Network},
journal = {Advanced Quantum Technologies},
volume = {8},
number = {10},
pages = {e2500224},
year = {2025},
doi = {https://doi.org/10.1002/qute.202500224},
url = {https://advanced.onlinelibrary.wiley.com/doi/abs/10.1002/qute.202500224}
}

@article{25shijieqknn,
title = {High-rate discretely-modulated continuous-variable quantum key distribution using quantum machine learning},
journal = {Chaos, Solitons \& Fractals},
volume = {196},
pages = {116331},
year = {2025},
issn = {0960-0779},
doi = {https://doi.org/10.1016/j.chaos.2025.116331},
url = {https://www.sciencedirect.com/science/article/pii/S0960077925003443},
author = {Qin Liao and Zhuoying Fei and Jieyu Liu and Anqi Huang and Lei Huang and Yijun Wang}
}

@article{06durr,
author = {D\"{u}rr, Christoph and Heiligman, Mark and HOyer, Peter and Mhalla, Mehdi},
title = {Quantum Query Complexity of Some Graph Problems},
journal = {SIAM Journal on Computing},
volume = {35},
number = {6},
pages = {1310-1328},
year = {2006},
doi = {10.1137/050644719},
URL = {https://doi.org/10.1137/050644719},
eprint = {https://doi.org/10.1137/050644719}
}

@Inbook{10supervise,
author="Tsoumakas, Grigorios
and Katakis, Ioannis
and Vlahavas, Ioannis",
editor="Maimon, Oded
and Rokach, Lior",
title="Mining Multi-label Data",
bookTitle="Data Mining and Knowledge Discovery Handbook",
year="2010",
publisher="Springer US",
address="Boston, MA",
pages="667--685",
isbn="978-0-387-09823-4",
doi="10.1007/978-0-387-09823-4_34",
url="https://doi.org/10.1007/978-0-387-09823-4_34"
}

@article{16dimension,
title = {A systematic review of multi-label feature selection and a new method based on label construction},
journal = {Neurocomputing},
volume = {180},
pages = {3-15},
year = {2016},
note = {Progress in Intelligent Systems Design},
issn = {0925-2312},
doi = {https://doi.org/10.1016/j.neucom.2015.07.118},
url = {https://www.sciencedirect.com/science/article/pii/S0925231215016197},
author = {Newton Spolaôr and Maria Carolina Monard and Grigorios Tsoumakas and Huei Diana Lee}
}

@article{21textclassify,
title = {Multi-label text classification via joint learning from label embedding and label correlation},
journal = {Neurocomputing},
volume = {460},
pages = {385-398},
year = {2021},
issn = {0925-2312},
doi = {https://doi.org/10.1016/j.neucom.2021.07.031},
url = {https://www.sciencedirect.com/science/article/pii/S0925231221010754},
author = {Huiting Liu and Geng Chen and Peipei Li and Peng Zhao and Xindong Wu}
}

@InProceedings{20gene,
author="Tarekegn, Adane
and Ricceri, Fulvio
and Costa, Giuseppe
and Ferracin, Elisa
and Giacobini, Mario",
editor="Hu, Ting
and Louren{\c{c}}o, Nuno
and Medvet, Eric
and Divina, Federico",
title="Detection of Frailty Using Genetic Programming",
booktitle="Genetic Programming",
year="2020",
publisher="Springer International Publishing",
address="Cham",
pages="228--243",
isbn="978-3-030-44094-7"
}

@ARTICLE{67knn,
  author={Cover, T. and Hart, P.},
  journal={IEEE Transactions on Information Theory}, 
  title={Nearest neighbor pattern classification}, 
  year={1967},
  volume={13},
  number={1},
  pages={21-27},
  doi={10.1109/TIT.1967.1053964}}

@inproceedings{96grover,
author = {Grover, Lov K.},
title = {A fast quantum mechanical algorithm for database search},
year = {1996},
isbn = {0897917855},
publisher = {Association for Computing Machinery},
address = {New York, NY, USA},
url = {https://doi.org/10.1145/237814.237866},
doi = {10.1145/237814.237866},
booktitle = {Proceedings of the Twenty-Eighth Annual ACM Symposium on Theory of Computing},
pages = {212–219},
numpages = {8},
location = {Philadelphia, Pennsylvania, USA},
series = {STOC '96}
}

@book{10compute, 
place={Cambridge}, 
title={Quantum Computation and Quantum Information: 10th Anniversary Edition}, 
publisher={Cambridge University Press}, 
author={Nielsen, Michael A. and Chuang, Isaac L.}, 
year={2010}}

@article{19pca,
author = {Yu, Chao-Hua and Gao, Fei and Lin, Song and Wang, Jingbo},
title = {Quantum data compression by principal component analysis},
journal = {Quantum Information Processing},
year = {2019},
volume = {18},
number = {8},
pages = {249},
month = {7},
issn = {1573-1332},
doi = {10.1007/s11128-019-2364-9},
url = {https://doi.org/10.1007/s11128-019-2364-9}
}

@misc{02qae,
   title={Quantum amplitude amplification and estimation},
   ISSN={0271-4132},
   url={http://dx.doi.org/10.1090/conm/305/05215},
   DOI={10.1090/conm/305/05215},
   journal={Quantum Computation and Information},
   publisher={American Mathematical Society},
   author={Brassard, Gilles and Høyer, Peter and Mosca, Michele and Tapp, Alain},
   year={2002},
   pages={53–74} }

@article{23correlation,
title = {Multi-label feature selection based on correlation label enhancement},
journal = {Information Sciences},
volume = {647},
pages = {119526},
year = {2023},
issn = {0020-0255},
doi = {https://doi.org/10.1016/j.ins.2023.119526},
url = {https://www.sciencedirect.com/science/article/pii/S0020025523011118},
author = {Zhuoxin He and Yaojin Lin and Chenxi Wang and Lei Guo and Weiping Ding}
}

@article{19categorization,
title = {Multi-label Arabic text categorization: A benchmark and baseline comparison of multi-label learning algorithms},
journal = {Information Processing \& Management},
volume = {56},
number = {1},
pages = {212-227},
year = {2019},
issn = {0306-4573},
doi = {https://doi.org/10.1016/j.ipm.2018.09.008},
url = {https://www.sciencedirect.com/science/article/pii/S0306457318300736},
author = {Bassam Al-Salemi and Masri Ayob and Graham Kendall and Shahrul Azman Mohd Noah}
}

@InProceedings{15tree,
author="Wang, Xiaoxue
and An, Shuang
and Shi, Hong
and Hu, Qinghua",
editor="Yao, Yiyu
and Hu, Qinghua
and Yu, Hong
and Grzymala-Busse, Jerzy W.",
title="Fuzzy Rough Decision Trees for Multi-label Classification",
booktitle="Rough Sets, Fuzzy Sets, Data Mining, and Granular Computing",
year="2015",
publisher="Springer International Publishing",
address="Cham",
pages="207--217",
isbn="978-3-319-25783-9"
}

@article{23kernel,
title = {Ensemble of kernel extreme learning machine based elimination optimization for multi-label classification},
journal = {Knowledge-Based Systems},
volume = {278},
pages = {110817},
year = {2023},
issn = {0950-7051},
doi = {https://doi.org/10.1016/j.knosys.2023.110817},
url = {https://www.sciencedirect.com/science/article/pii/S0950705123005671},
author = {Qingshuo Zhang and Eric C.C. Tsang and Qiang He and Yanting Guo}
}

@inproceedings{01gene-kernel,
author = {Elisseeff, Andr\'{e} and Weston, Jason},
title = {A kernel method for multi-labelled classification},
year = {2001},
publisher = {MIT Press},
address = {Cambridge, MA, USA},
booktitle = {Proceedings of the 15th International Conference on Neural Information Processing Systems: Natural and Synthetic},
pages = {681–687},
numpages = {7},
location = {Vancouver, British Columbia, Canada},
series = {NIPS'01}
}

@article{10pca,
author = {Abdi, Herv\'{e} and Williams, Lynne J.},
title = {Principal component analysis},
year = {2010},
issue_date = {July 2010},
publisher = {John Wiley \\& Sons, Inc.},
address = {USA},
volume = {2},
number = {4},
issn = {1939-5108},
url = {https://doi.org/10.1002/wics.101},
doi = {10.1002/wics.101},
journal = {WIREs Comput. Stat.},
month = jul,
pages = {433–459},
numpages = {27}
}

@article{19kernel,
author = {Havlíček, Vojtěch and Córcoles, Antonio D. and Temme, Kristan and Harrow, Aram W. and Kandala, Abhinav and Chow, Jerry M. and Gambetta, Jay M.},
title = {Supervised learning with quantum-enhanced feature spaces},
journal = {Nature},
volume = {567},
number = {7747},
pages = {209--212},
year = {2019},
month = mar,
doi = {10.1038/s41586-019-0980-2},
url = {https://doi.org/10.1038/s41586-019-0980-2},
issn = {1476-4687}
}

@article{17qml,
  author  = {Biamonte, Jacob and Wittek, Peter and Pancotti, Nicola and Rebentrost, Patrick and Wiebe, Nathan and Lloyd, Seth},
  title   = {Quantum machine learning},
  journal = {Nature},
  volume  = {549},
  number  = {7671},
  pages   = {195--202},
  year    = {2017},
  month   = sep,
  doi     = {10.1038/nature23474},
  url     = {https://doi.org/10.1038/nature23474},
  issn    = {1476-4687}
}

@article{18imageknn,
author = {Dang, Yijie and Jiang, Nan and Hu, Hao and Ji, Zhuoxiao and Zhang, Wenyin},
title = {Image classification based on quantum K-Nearest-Neighbor algorithm},
year = {2018},
issue_date = {September 2018},
publisher = {Kluwer Academic Publishers},
address = {USA},
volume = {17},
number = {9},
issn = {1570-0755},
url = {https://doi.org/10.1007/s11128-018-2004-9},
doi = {10.1007/s11128-018-2004-9},
journal = {Quantum Information Processing},
month = sep,
pages = {1–18},
numpages = {18}
}

@article{20data-encode,
  title = {Robust data encodings for quantum classifiers},
  author = {LaRose, Ryan and Coyle, Brian},
  journal = {Phys. Rev. A},
  volume = {102},
  issue = {3},
  pages = {032420},
  numpages = {24},
  year = {2020},
  month = {Sep},
  publisher = {American Physical Society},
  doi = {10.1103/PhysRevA.102.032420},
  url = {https://link.aps.org/doi/10.1103/PhysRevA.102.032420}
}

@Inbook{18complexity,
author="Schuld, Maria
and Petruccione, Francesco",
title="Quantum Advantages",
bookTitle="Supervised Learning with Quantum Computers",
year="2018",
publisher="Springer International Publishing",
address="Cham",
pages="127--137",
isbn="978-3-319-96424-9",
doi="10.1007/978-3-319-96424-9_4",
url="https://doi.org/10.1007/978-3-319-96424-9_4"
}

@article{22weight-mlknn,
title = {A weighted ML-KNN based on discernibility of attributes to heterogeneous sample pairs},
journal = {Information Processing \& Management},
volume = {59},
number = {5},
pages = {103053},
year = {2022},
issn = {0306-4573},
doi = {https://doi.org/10.1016/j.ipm.2022.103053},
url = {https://www.sciencedirect.com/science/article/pii/S0306457322001571},
author = {Xin Wen and Deyu Li and Chao Zhang and Yanhui Zhai}
}

@article{21textclassify2,
title = {A novel reasoning mechanism for multi-label text classification},
journal = {Information Processing \& Management},
volume = {58},
number = {2},
pages = {102441},
year = {2021},
issn = {0306-4573},
doi = {https://doi.org/10.1016/j.ipm.2020.102441},
url = {https://www.sciencedirect.com/science/article/pii/S0306457320309341},
author = {Ran Wang and Robert Ridley and Xi’ao Su and Weiguang Qu and Xinyu Dai}
}

@article{23weighted,
author = {Wu, Hongxin and Han, Meng and Chen, Zhiqiang and Li, Muhang and Zhang, Xilong},
title = {A Weighted Ensemble Classification Algorithm Based on Nearest Neighbors for Multi-Label Data Stream},
year = {2023},
issue_date = {June 2023},
publisher = {Association for Computing Machinery},
address = {New York, NY, USA},
volume = {17},
number = {5},
issn = {1556-4681},
url = {https://doi.org/10.1145/3570960},
doi = {10.1145/3570960},
month = feb,
articleno = {72},
numpages = {21}
}

@article{24Self-Adjusting,
author = {Nicola, Victor Gomes De Oliveira Martins and Delgado, Karina Valdivia and Lauretto, Marcelo de Souza},
title = {Imbalance-Robust Multi-Label Self-Adjusting kNN},
year = {2024},
issue_date = {September 2024},
publisher = {Association for Computing Machinery},
address = {New York, NY, USA},
volume = {18},
number = {8},
issn = {1556-4681},
url = {https://doi.org/10.1145/3663575},
doi = {10.1145/3663575},
journal = {ACM Trans. Knowl. Discov. Data},
month = jul,
articleno = {190},
numpages = {30}
}

@article{97diffusion,
  title = {Quantum Mechanics Helps in Searching for a Needle in a Haystack},
  author = {Grover, Lov K.},
  journal = {Phys. Rev. Lett.},
  volume = {79},
  issue = {2},
  pages = {325--328},
  numpages = {0},
  year = {1997},
  month = {Jul},
  publisher = {American Physical Society},
  doi = {10.1103/PhysRevLett.79.325},
  url = {https://link.aps.org/doi/10.1103/PhysRevLett.79.325}
}

@misc{14superposition-state,
      title={Quantum Algorithms for Nearest-Neighbor Methods for Supervised and Unsupervised Learning}, 
      author={Nathan Wiebe and Ashish Kapoor and Krysta Svore},
      year={2014},
      eprint={1401.2142},
      archivePrefix={arXiv},
      primaryClass={quant-ph},
      url={https://arxiv.org/abs/1401.2142}, 
}

@article{26QADS,
title = {Detecting quantum hacking attacks for continuous-variable quantum key distribution using quantum neural network},
journal = {Chaos, Solitons \& Fractals},
volume = {202},
pages = {117467},
year = {2026},
issn = {0960-0779},
doi = {https://doi.org/10.1016/j.chaos.2025.117467},
url = {https://www.sciencedirect.com/science/article/pii/S0960077925014808},
author = {Junhao Li and Yiyu Mao and Qin Liao and Yan Ding and Zhuo Tang and Kenli Li}
}

@article{21qknn,
title = {Quantum K-Nearest-Neighbor Image Classification Algorithm Based on K-L Transform},
journal = {International Journal of Theoretical Physics},
volume = {60},
year = {2021},
issn = {1572-9575},
doi = {https://doi.org/10.1007/s10773-021-04747-7},
author = {Zhou and NR. and Liu and XX. and Chen and YL}
}

@article{22cswap,
title = {New approach of KNN Algorithm in quantum computing based on new design of quantum circuits}, 
volume={46}, 
url={https://www.informatica.si/index.php/informatica/article/view/3608}, 
DOI={10.31449/inf.v46i5.3608},
number={5},
journal={Informatica},
year={2022},
month={Apr.},
author={Vu Tuan Hai and Phan Hoang Chuong and Pham The Bao}
}

@book{94sorting,
  title={Introduction to algorithms},
  author={Leiserson, Charles Eric and Rivest, Ronald L and Cormen, Thomas H and Stein, Clifford},
  volume={3},
  year={1994},
  publisher={MIT press Cambridge, MA, USA}
}

\end{document}